\documentclass[%
reprint,
unsortedaddress,
 amsmath,amssymb,
 aps,
]{revtex4-2}

\usepackage[utf8]{inputenc}
\usepackage{graphicx}
\usepackage{dcolumn}
\usepackage{multirow}
\usepackage{bm}
\usepackage{mathrsfs}
\usepackage{hyperref}
\hypersetup{
  colorlinks = true,
  urlcolor = blue,
  linkcolor = blue,
  citecolor = green,
  filecolor = magenta,
}

\newcommand{\ud}{\mathrm{d}}
\newcommand{\pd}{\partial}

\begin{document}


\title{Lensing rates of gravitational wave signals displaying beat patterns detectable by DECIGO and B-DECIGO}

\author{Shaoqi Hou}
\affiliation{School of Physics and Technology, Wuhan University, Wuhan, Hubei 430072, China}
\author{Pengbo Li}
\affiliation{School of Physics and Technology, Wuhan University, Wuhan, Hubei 430072, China}
\author{Hai Yu}
\affiliation{Department of Astronomy, School of Physics and Astronomy, Shanghai Jiao Tong University, Shanghai, 200240, China}
\author{Marek Biesiada}
\affiliation{Department of Astronomy, Beijing Normal University, Beijing 100875, China}
\affiliation{National Centre for Nuclear Research, Pasteura 7, 02-093 Warsaw, Poland}
\author{Xi-Long Fan}
\affiliation{School of Physics and Technology, Wuhan University, Wuhan, Hubei 430072, China}
\author{Seiji Kawamura}
\affiliation{Department of Physics, Nagoya University, Nagoya, Aichi 464-8602, Japan}
\author{Zong-Hong Zhu}
\email{zhuzh@whu.edu.cn}
\affiliation{School of Physics and Technology, Wuhan University, Wuhan, Hubei 430072, China}
\affiliation{Department of Astronomy, Beijing Normal University, Beijing 100875,  China}

\date{\today}

\begin{abstract}
The coherent nature of gravitational wave emanating from a compact binary system makes it possible to detect some interference patterns in two (or more) signals registered simultaneously by the detector.
Gravitational lensing effect can be used to bend trajectories of gravitational waves, which might reach the detector at the same time. 
Once this happens, a beat pattern may form, and can be used to obtain the luminosity distance of the source, the lens mass, and cosmological parameters such as the Hubble constant.
Crucial question is how many such kind of events could be detected.
In this work, we study this issue for the future space-borne detectors: DECIGO and its downscale version, B-DECIGO.
It is found out that there can be a few tens to a few hundreds of lensed gravitational wave events with the beat pattern observed by DECIGO and B-DECIGO per year, depending on the evolution scenario leading to the formation of double compact objects.
In particular, black hole-black hole binaries are dominating population of lensed sources in which beat patterns may reveal. However, DECIGO could also register a considerable amount of lensed signals from binary neutron stars, which might be accompanied by electromagnetic counterparts.   
Our results suggest that, in the future, lensed gravitational wave signal with the beat pattern could play an important role in cosmology and astrophysics.
\end{abstract}

\maketitle


\section{Introduction}\label{eq-intro}

Einstein's prediction of gravitational waves (GWs) \cite{Einstein:1916cc,*Einstein:1918btx} has been verified by the detection of GWs by LIGO/Virgo Collaborations \cite{Abbott:2016blz,Abbott:2016nmj,Abbott:2017vtc,Abbott:2017oio,TheLIGOScientific:2017qsa,Abbott:2017gyy,LIGOScientific:2018mvr,Abbott:2020uma,LIGOScientific:2020stg,Abbott:2020khf,Abbott:2020tfl,Abbott:2020mjq,Abbott:2020niy}.
These observations marked the new era of GW astronomy and multimessenger astrophysics \cite{TheLIGOScientific:2017qsa,Goldstein:2017mmi,*Savchenko:2017ffs}.
Together with its electromagnetic counterpart, the GW could shed new light on cosmology, since its source can be used as the standard siren to accurately measure the luminosity distance \cite{Schutz:1986gp,*Holz:2005df}. 
It is also interesting to know that it may not be necessary to rely on the electromagnetic counterpart to determine the redshift of the source as discussed in Refs.~\cite{Seto:2001qf,*Nishizawa:2011eq,*Bonvin:2016qxr,Messenger:2011gi,*Messenger:2013fya,Hou:2019jhu,Ding_2019}.
The GW can also serve as a probe into fundamental physics \cite{Barausse:2020rsu}, such as  the nature of gravity \cite{Nishizawa:2009bf,*Will:2014kxa,*Hou:2017bqj,*Hou:2018djz,*Gong:2018ybk,*Gong:2018cgj,*Hou:2018mey,*Hohmann:2019nat,*Liu:2019cxm} and spacetime \cite{Christodoulou1991,*Flanagan:2015pxa,*Du:2016hww,*Hou:2020tnd,*Hou:2020wbo,*Hou:2020xme}.

The GW travels at the speed of light in vacuum $c$ (i.e. along null geodesic) as predicted by general relativity \cite{Misner:1974qy}. If there is a massive enough object near its path, the trajectory of the GW is bent due to the curvature of spacetime produced by this object.  
This is the gravitational lensing effect \cite{gravlens1992,Lawrence1971nc,*Lawrence:1971hx}.
Although the light can also be gravitationally lensed, one should understand that there are several differences between the lensing of the light and that of the GW.
First, GWs usually have much longer wavelengths than the light.
Second, the GW produced by a compact binary system is nearly monochromatic, so it is coherent; in contrast, the light emitted by a star is incoherent.
Because of the long wavelengths of GWs, the lens should be very massive in order for the geometric optics to be applicable. 
For example, the mass of the lens should be greater than $10^4M_\odot$ for the ground-based detectors, which are operating in the frequency range $1\sim10^4$ Hz; 
for LISA (frequency range $10^{-4}\sim0.1$ Hz), the lens should be $10^8M_\odot$, at least \cite{ArnaudVarvella:2003va,Meena:2019ate}. 
Although we will refer to the wave nature of GWs, we focus on the geometric optics regime.
In this regime, the lensed GW has magnified amplitude and its polarization plane gets rotated \cite{Hou:2019wdg}.
However, the deflection angle is very small, so the rotation of the polarization plane can be ignored.

As an effect of the gravitational lensing, there can be multiple paths along which  GWs reach the detector.
The GWs along different paths experience distinct gravtitational time dilation, and the lengths of the paths are not the same, so there exist time delays $\Delta t$ between them \cite{gravlens1992}. 
If it happens that, during some time window, the interferometer simultaneously detects the GWs coming from the same source along different trajectories, interference patterns may occur \cite{Yamamoto:2005ea,Hou:2019dcm}.
During the inspiral phase, frequency of the GW evolves very slowly. Consequently, there might be a small frequency difference $\Delta f$ between the GWs coming from the lensed images of the source. 
Therefore, if $\Delta f$ is small enough the interference results in a beat pattern, which can be used to extract useful information (e.g., lensing time delay $\Delta t$ and the magnifications) and further to measure the true luminosity distance of the source, the lens mass, and cosmological parameters as discussed in Ref.~\cite{Hou:2019dcm}. 
In typical cases of galaxy lensing, time delay $\Delta t$ might be of order of a few days to a few months. 
Therefore, it is highly impossible for a ground-based detector operating at high frequencies to simultaneously observe the lensed GWs traveling along different paths.
This is because the GW from the final merger phase detectable in the frequency range of the ground-based detector lasts for a few hundred seconds at most. 
However, there is no difficulty for the space-borne detector sensitive to low GW frequencies to observe the beat pattern.  
Therefore, it is very interesting to study the prospects of future space-borne GW detectors to register the beat patterns from lensed GW signals.

As a starting point, one should estimate how many lensed GW events exhibiting the beat pattern to be detected by the space-borne detector per year.
If the detection rate is large enough, it would definitely be important to study such phenomenon further. 
In the past, Sereno \textit{et al.} calculated the number of lensed GW events that might be observed by LISA \cite{Sereno:2010dr}. 
They found out that there can be at most 4 lensed GW events in a 5-year mission. 
Since they did not specialize the particular events with the beat pattern, one expects that those with the beat pattern should be fewer than 4. 
Nevertheless, it is worth to note that even with such a low detection rate, some interesting constraints on cosmological parameters can still be derived according to Ref.~\cite{Sereno:2011ty}.
One expects that with a higher detection rate, the constraints can be improved.
The low detection rate is related to the fact that the number of the massive black hole binaries, the main targets of LISA, is only on the order of $10^3$ \cite{Sesana:2008ur,Klein:2015hvg}. 
On the contrary, there are many more less massive compact binaries, whose merger rate is larger by a few orders of magnitude \cite{LIGOScientific:2018mvr,Gerosa:2019dbe,Boco_2019}.
These mergers might be easier detected by a second space-borne interferometer, DECi-hertz Interferometer Gravitational wave Observatory (DECIGO) \cite{Seto:2001qf,Kawamura:2011zz,Kawamura:2020pcg}. 

DECIGO is a planed Japanese space-borne interferometer, which is sensitive to GWs at mHz to 100 Hz.
Since its sensitive band is higher than that of LISA, DECIGO  is capable of observing GWs from much less massive binaries.
Moreover, some of these binaries would also be targets of ground-based detectors, such as LIGO/Virgo and Einstein Telescope (ET) \cite{Punturo:2010zza}.
Therefore, DECIGO and ground-based detectors can observe some merging binary systems jointly (but at different times, of course) to make the multiband GW astronomy possible \cite{Sesana:2016ljz}. 
B-DECIGO, which is a downscale version of DECIGO, will also be operating in the similar frequency band, but will be less sensitive \cite{Sato_2017,decigo2019}.
In this work, we discuss the detection rates of the lensed GW events exhibiting the beat pattern observable by (B-)DECIGO. 

The predictions of the GW lensing rate have been formulated by several authors for different interferometers. 
As mentioned above, Sereno \emph{et al.} estimated how many lensed GW events can be detected by LISA \cite{Sereno:2010dr}.
For ground-based detectors, Ref.~\cite{Li:2018prc} predicted that there would be only 1 lensed GW event per year for aLIGO at its design sensitivity, but ET can detect about 80 events.
Refs.~\cite{Piorkowska:2013eww,Biesiada:2014kwa,Ding:2015uha,Oguri:2018muv,Yang:2019jhw} all discussed the lensing rate for ET, and concluded that there are $50 \sim 100$ events per year. 
Recently, Ref.~\cite{Piorkowska-Kurpas:2020mst} predicted the lensing rates for (B-)DECIGO, but the possibility of the beat pattern was not considered.
This work will fill in the gap.

Gravitational lensing has many  applications other than those discussed in Ref.~\cite{Hou:2019dcm}. 
For instance, one can detect dark matter \cite{Cutler:2009qv,Camera:2013xfa,Congedo:2018wfn,Jung:2017flg,Liao_2018}, constrain the speed of light \cite{Fan:2016swi,Collett:2016dey}, determine the cosmological constant \cite{Sereno:2010dr,Sereno:2011ty,Liao:2017ioi}, examine the wave nature of GWs \cite{Dai:2018enj,Liao:2019aqq,Sun:2019ztn}, and localize the host galaxies of strong lensed GWs \cite{Yu2020MNRAS.497..204Y,Hannuksela2020MNRAS.tmp.2493H} using gravitational lensing.
Although no gravitational lensing signals have been detected in the observed GW events, the advent of more sensitive GW detectors might make it possible soon \cite{Hannuksela:2019kle}.

This work is organized in the following way.
We will start with a brief review of the formation of the beat pattern due to the lensing effect in Sec.~\ref{sec-beat}.
Then, (B-)DECIGO will be introduced in Sec.~\ref{sec-det}.
Section~\ref{sec-tau} discusses the lens model and how to calculate the optical depth, and the lensing rates are computed in Sec.~\ref{sec-br} mainly following Ref.~\cite{Ding:2015uha}. 
In the end, there is a short conclusion in Sec.~\ref{sec-con}.
We choose a units such that $c=1$.

\section{The beat pattern}
\label{sec-beat}

In this section, we shall briefly review the formation of the beat pattern due to the gravitational lensing effect of GWs.
For more detail, please refer to Ref.~\cite{Hou:2019dcm}.

\begin{figure}[h]
    \centering
    \includegraphics[width=0.45\textwidth]{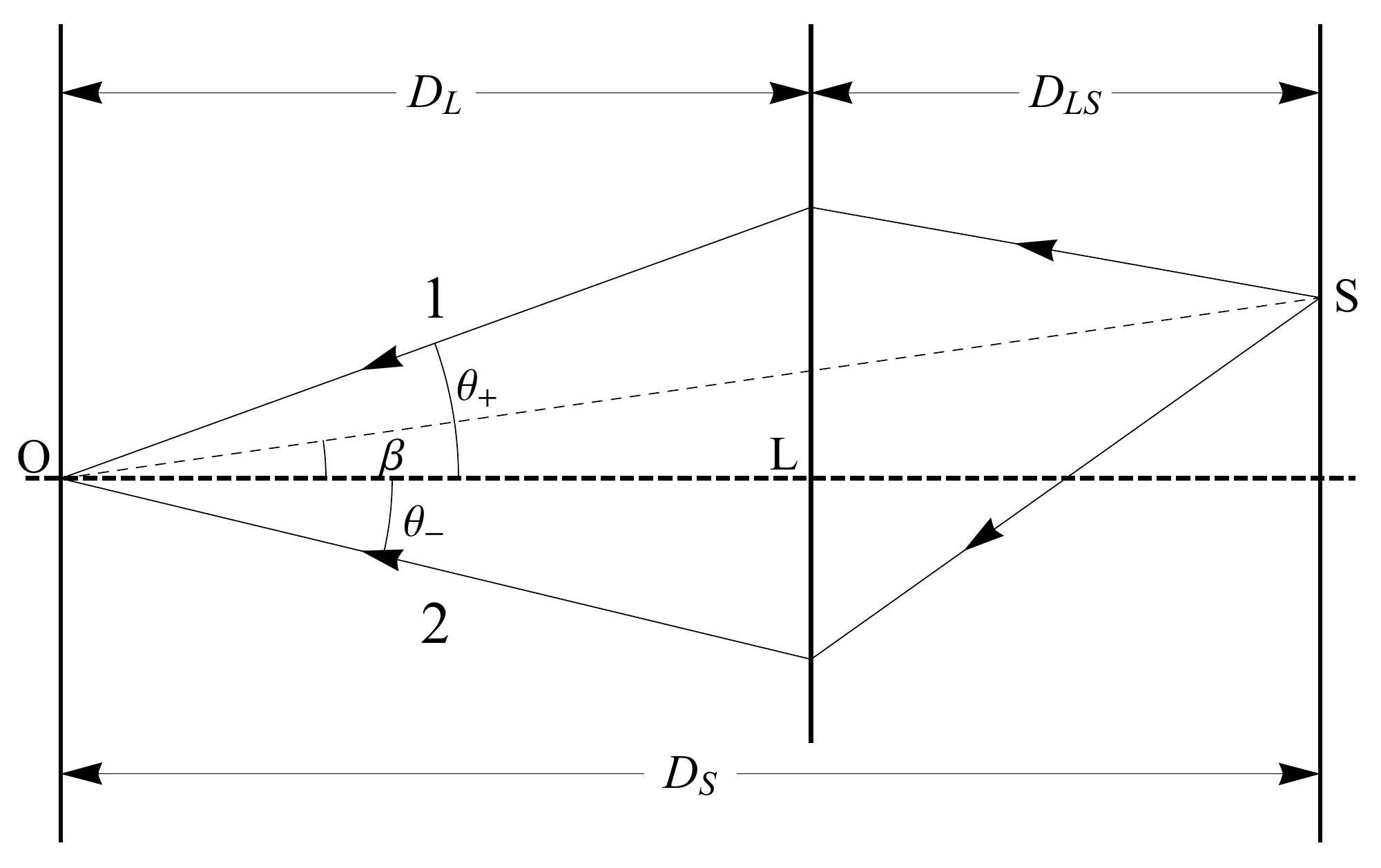}
    \caption{The geometry of a lens.
    Two GW rays, 1 and 2, originating from the source S, travel along two trajectories and change their directions near the lens L.
    Eventually, they arrive at the detector at O. 
    Vertical lines represent the observer, lens and source planes, from the left to the right. 
    Thick dashed line is the optical axis, and the thin dashed line would be the viewing direction if there were no lens. 
    The angle between the two dashed lines is called the misalignment angle $\beta$.
    The GW rays form the angles $\theta_+$ and $\theta_-$ with the optical axis at the observer.
    The distances $D_\text{S}, D_\text{L},$ and $D_\text{LS}$ are angular diameter distances.}
    \label{fig-geo}
\end{figure}

We will assume that the lens is described by a singular isothermal sphere (SIS), which models the early type galaxies, because they contribute to the strong lensing probability dominantly \cite{Turner:1984ch,*Moeller:2006cu}.
The line-of-sight velocity dispersion $\sigma$ of stars in the galaxy characterizes the lensing effect. 
As shown in Fig.~\ref{fig-geo}, the GW produced by the source S can travel in two paths, labeled by 1 and 2, to arrive at the observer O due to the presence of a lens L. 
$\beta$ is the misalignment angle between the optical axis OL and the would-be viewing direction OS if there were no lens.
Deflected rays form two angles, $\theta_\pm$, with OL at the observer, which are given by  \cite{gravlens1992}
\begin{equation}
    \label{eq-th-pm}
 \theta_\pm=\beta\pm\theta_\text{E},
\end{equation}
where $\theta_\text{E}=4\pi\sigma^2D_\text{LS}/D_\text{S}$ is the angular Einstein radius, and $D_\text{LS}$ and $D_\text{S}$ are the angular diameter distances indicated in the Fig.~\ref{fig-geo}. 
In further considerations concerning merger rates we will assume flat $\Lambda$CDM model, in which 
 the angular diameter distance $D_\text{A}(z)$ between the Earth and a celestial object at the redshift $z$ is \cite{Weinberg:2008zzc}
\begin{equation}
  \label{eq-daw}
  D_\text{A}(z) = \frac{1}{H_0 (1+z)} \int_0^z \frac{dz'}{h(z')}
\end{equation}
where $H_0$ is the Hubble constant, $ h(z)=[\Omega_m (1+z)^3 + \Omega_{\Lambda}]^{1/2}$ is the dimensionless expansion rate. In order to comply with population synthesis model used further in this paper, we assume $\Omega_m = 0.3$ and $H_0 = 70 \; \text{km s}^{-1} \text{Mpc}^{-1}$.
Therefore, in our shorthand notation: $D_\text{L}=D_\text{A}(z_\text{L})$, and $D_\text{S}=D_\text{A}(z_\text{S})$ with $z_\text{L}$ and $z_\text{S}$ the redshifts of the lens and the source, respectively.
$D_\text{LS} = \frac{1}{H_0 (1+z_\text{S})} \int_{z_\text{L}}^{z_\text{S}} \frac{dz'}{h(z')}$ is the angular diameter distance between the lens and the source. 

Lensed GW signals are magnified, and the magnification factors of the GW amplitudes are given by 
\begin{equation}\label{eq-mu-sis}
  \mu_\pm=\sqrt{\left|\frac{\theta_\pm/\theta_\text{E}}{|\theta_\pm/\theta_\text{E}|-1}\right|}.
\end{equation}
Finally, the GW rays travel along paths of different lengths, and they experience different time dilation due to the gravitational potential of the lens, so they arrive at the observer at different times. 
The time delay is 
\begin{equation}\label{eq-tdv-sis}
  \Delta t  =32\pi^2\sigma^4(1+z_\text{L})\frac{D_\text{L}D_\text{LS}}{D_\text{S}}\frac{\beta}{\theta_\text{E}}.
\end{equation}
One can see that $\sigma$, appearing in Eqs.~\eqref{eq-th-pm}, \eqref{eq-mu-sis}, and \eqref{eq-tdv-sis}, indeed characterizes the lensing effect of the SIS model.

The time delay $\Delta t$ typically ranges from a few days to a few months. 
For example, one can assume that the GW source is at $z_\text{S}=2$, and the lens at $z_\text{L}=1$. 
Such value of $z_\text{S}$ is suggested by the fact that the redshift distribution of detectable neutron star-neutron star mergers (NS-NS) is maximal  near $z=2$, the redshift of black hole-black hole mergers (BH-BH) peaks around $z=4$ \cite{Yang:2019jhw} and the lensing probability is maximal roughly for a lens half-way between the source and observer.
Taking $\sigma$ as a characteristic velocity dispersion $\sigma_*=161\pm5\text{ km/s}$ \cite{Choi:2006qg}, then $2.03$ weeks $<\Delta t<$ 1.18 months for $0.1\text{ arcsecond}<\beta<0.25$ arcsecond. 
Note that $\beta<\theta_\text{E}\approx0.27$ arcsecond in order that the interferometer can ``hear'' two GWs.

One can reasonably expect that $\Delta t$ is much longer than the duration of GW signal observed by ground-based interferometers. 
Of course, one may imagine a case where $\beta$ is very small, of the order of $10^{-5}-10^{-7}$ arcsecond, such that $\Delta t$ is of order of a few seconds, and the beat pattern forms, as displayed in Fig.~2 in Ref.~\cite{Hou:2019dcm}.
But the probability for such case is extremely low. 
Therefore, it is very unlikely to use ground-based interferometers to observe the interference patterns in GW events lensed by a SIS. 
However, the signals observed by LISA and (B-)DECIGO usually last for  several months or even years. So, there is no difficulty for them to simultaneously detect two lensed GW signals in some time window.
These signals would interfere with each other and form an interference pattern in the time domain. 
If the strains for the lensed GWs are $h_1(t)$ and $h_2(t)$, the total strain is simply
\begin{equation}
    \label{eq-ht}
    h(t)=h_1(t)+h_2(t).
\end{equation}
Suppose the frequencies of $h_1$ and $h_2$ are $f_1$ and $f_2$, respectively. 
Without the loss of generality, let $f_1>f_2$, i.e., we assume $h_1$ arrives earlier than $h_2$.
Their difference $\Delta f=f_1-f_2$ is much smaller than both $f_1$ and $f_2$ in the inspiral phase, due to slow evolution of the GW frequency during the lensing time delay.  
Therefore, the beat pattern could show up in the inspiral phase, as discussed in Ref.~\cite{Hou:2019dcm}.
As the binary system evolves, the GW frequency increases, so the beat pattern has  a smaller and smaller period. 
Eventually, the beat pattern disappears, and a generic interference pattern is left.

Taking into account the orbital motion of the space-borne interferometer, the beat pattern would have more complicated behavior than that described above. 
So one may want to consider the cases with small enough $\Delta t$ such that the impact on the beat pattern due to  the changing orientation of the constellation plane is small enough, and the analysis is easier. 
Of course, $\Delta t$ should not be too small; otherwise, the probability for such events would be negligible again. 
So in this work, we would like to mainly consider the lensing events with $\Delta t=1$ month.

The Fourier transformation of the strain is used to calculate the signal-to-noise ratio (SNR). 
Let $h_1(t)$ be Fourier transformed to 
\begin{equation}
    \label{eq-fh1}
  \begin{split}
  \tilde{h}_1(f)&=\mu_+\int_{-\infty}^\infty h_u(t)e^{i2\pi ft}\ud f\\
    =&\mu_+\tilde h_u(f),
  \end{split}
\end{equation} 
where $h_u(t)$ would be the strain if there were no lens, and $\tilde h_u(f)$ is its Fourier transform. 
Then the frequency domain waveform $\tilde h_2(f)$ for $h_2(t)=\frac{\mu_-}{\mu_+}h_1(t-\Delta t)$ is 
\begin{equation}
  \label{eq-h2f}
  \tilde h_2(f)=e^{i2\pi f\Delta t}\mu_-\tilde h_u(f).
\end{equation}
So the amplitude of the total waveform is 
\begin{equation}
  \label{eq-hf}
  |\tilde h(f)|=\sqrt{\mu_+^2+\mu_-^2+2\mu_+\mu_-\cos(2\pi f\Delta t)}|\tilde h_u(f)|.
\end{equation}
This suggests that in the frequency domain, the amplitude of the total waveform is also oscillating with a ``period'' $1/\Delta t$.

\section{DECIGO and B-DECIGO}
\label{sec-det}

In this work, we estimate the lensing rate of lensed GW events with beat patterns detectable by (B-)DECIGO, so this section briefly reviews the detector characteristics. 
DECIGO is supposed to have a configuration of four clusters of spacecrafts.
Each cluster would consists of three drag-free satellites, separated from each other by 1000 km and forming an equilateral triangle.
All four clusters would orbit around the Sun with a period of 1 yr.
DECIGO was originally proposed in Ref.~\cite{Seto:2001qf}.
Over the following years, it evolved somehow, and now, its current objectives and design can be found in Refs.~\cite{decigo2019,Kawamura:2020pcg}.

According to Yagi \& Seto \cite{Yagi:2011wg}, a triangular cluster is equivalent to two L-shaped interferometers rotated by $45^\circ$ with uncorrelated noise.
The noise spectrum for a single effective L-shape DECIGO is 
\begin{equation}
    \label{eq-dec-sh}
    \begin{split}
S_h(f) =& 10^{-48} \times \Bigg[ 7.05 \left( 1 + \frac{f^2}{f_p^2} \right) + 4.80\times 10^{-3}\times   \\ 
 & \frac{f^{-4}}{1 + \left( f/f_p \right)^2} + 5.33 \times 10^{-4} f^{-4} \Bigg]\;\; \text{Hz}^{-1},          
    \end{split}
\end{equation}
where $f_p = 7.36$ Hz. 

\begin{figure}[h]
    \centering
    \includegraphics[width=0.45\textwidth]{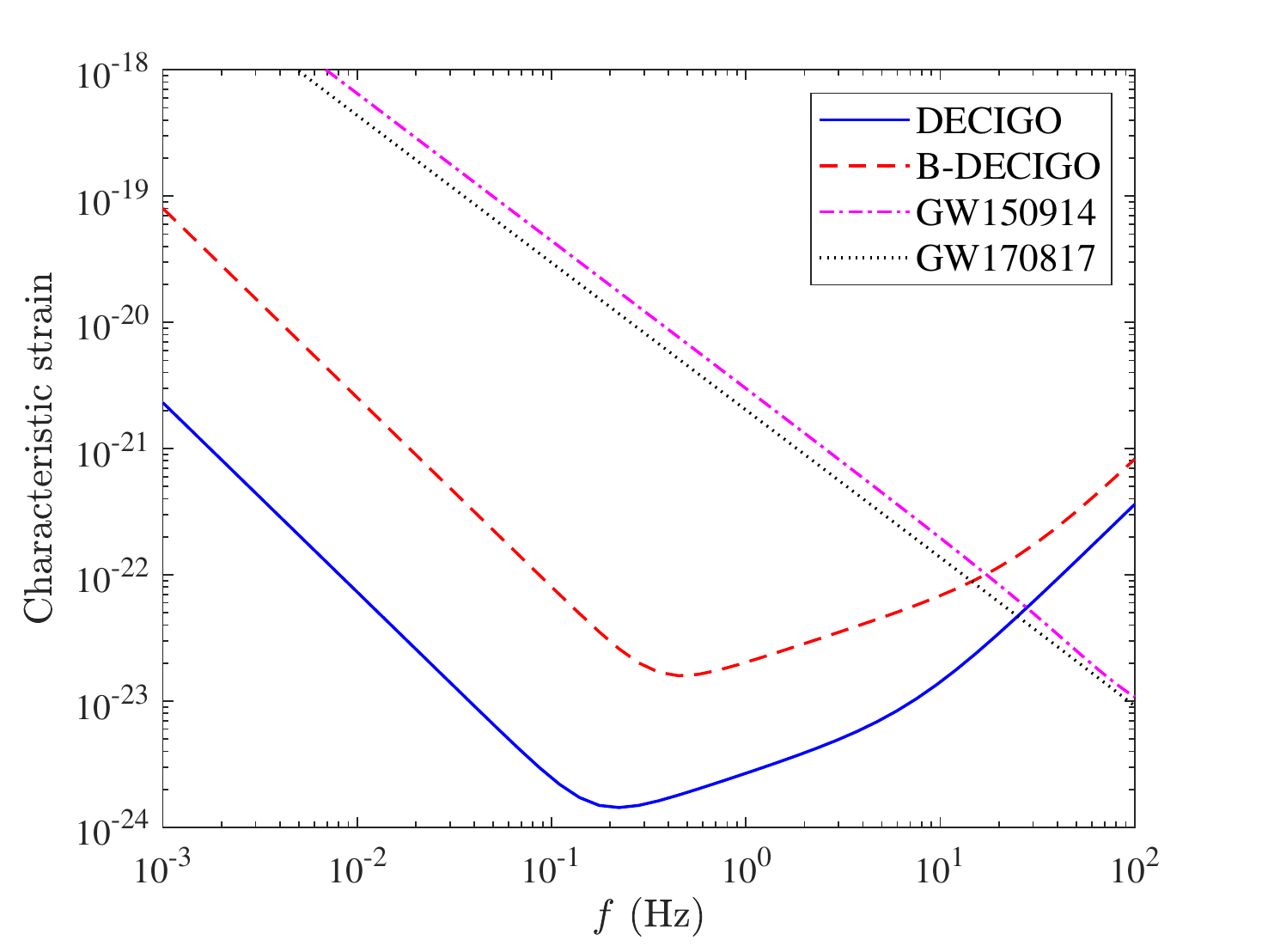}
    \caption{Characteristic strains of for DECIGO (solid blue curve) and B-DECIGO (dashed red curve). 
    The dot-dashed and the dotted curves are signals of GW150914 and GW170817, respectively.}
    \label{fig-chrstr}
\end{figure}

B-DECIGO  has only one cluster of spacecrafts.
The distance between spacecrafts is also smaller, which is 100 km. 
Its sensitivity is of course lower and can be described by the following effective noise spectrum \cite{Nakamura:2016hna,*Isoyama:2018rjb},
\begin{equation} 
    \label{eq-bdec-sh}
    \begin{split}
S_h(f) =& 10^{-46} \times [ 4.040 + 6.399 \times 10^{-2} f^{-4}   \\
& + 6.399 \times 10^{-3} f^{2} ]\;\; \text{Hz}^{-1}.
    \end{split}
\end{equation}
Fig.~\ref{fig-chrstr} shows the characteristic strains $\sqrt{fS_h}$ for these two detectors. 
For comparison, the signals of GW150914 and GW170817 are also plotted, using PyCBC \cite{alex_nitz_2019_2643618s}.
Although not shown in this figure, both of the two signals will later end with a  merger well beyond the sensitivity bands of (B-)DECIGO, but accessible to LIGO/Virgo and definitely also to next generation of ground-based detectors. 

\begin{figure}[h]
    \centering
    \includegraphics[width=.5\textwidth]{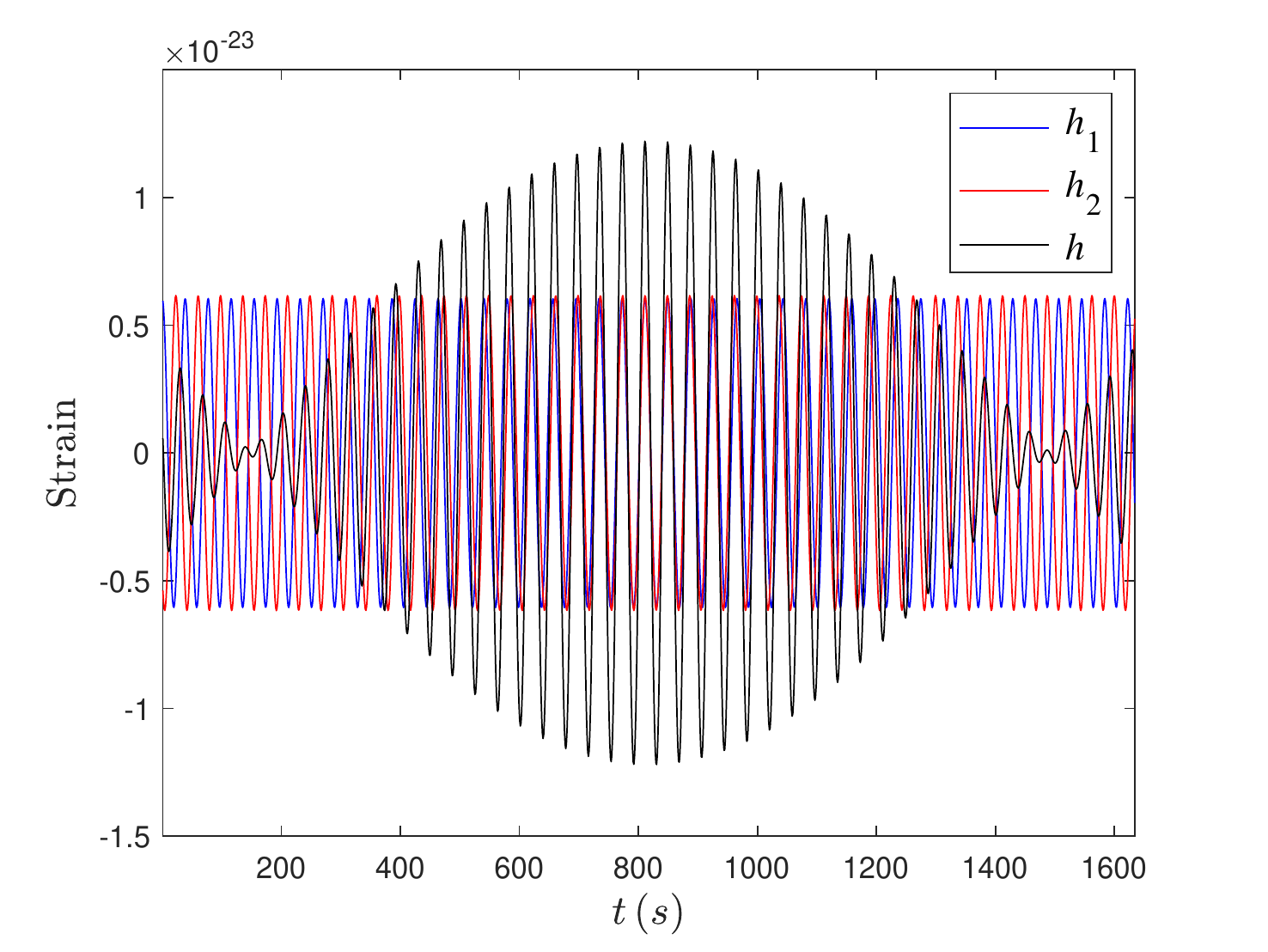}
    \caption{A schematic diagram showing the time domain waveform of the beat pattern. 
    The binary system is assumed to be GW150914, and the GW is gravitationally lensed by a suitable SIS lens. 
    The time delay is assumed to be 1 month.
    The blue and the red curves are for the strains of the GW rays propagating in two different trajectories, and the black curve is for the interfered wave.
    }
    \label{fig-bt}
\end{figure}

As one can see, both GW150914 and GW170817 would be detected by (B-)DECIGO in their inspiral phase. 
So if GW150914, for instance, were gravitationally lensed with a one-month time delay, then one may observe the beat pattern shown in Fig.~\ref{fig-bt}.
This figure displays the beat pattern (the black curve) formed due to the interference between two GW rays (the blue and the red curves) traveling in different paths because of a suitable SIS lens.
Here, for the purpose of demonstration, we only consider the quadruple contribution to the waveform, and assume the GWs incident the detector nearly perpendicularly and the inclination angle is zero.

Once one knows the Fourier transform $\tilde h(f)$ of a signal $h(t)$, one can calculate the SNR $\rho_\text{obs}$ of it \cite{Mirshekari:2011yq},
\begin{equation}
    \label{eq-def-snr}
    \rho_\text{obs}^2=4\int_0^\infty\frac{|\tilde h(f)|^2}{S_h(f)}\ud f.
\end{equation}
If $\rho_\text{obs}>\rho_\text{th}$, a threshold SNR, one may claim a detection of the GW.
This condition may not necessarily mean one can easily extract some useful information from the beat pattern. 
For that purpose, one expects that the higher the SNR is, the easier the extraction can be done. 
However, we will not determine the least SNR for an accurate extraction in this work, in spite of its importance.

\section{The lens model and the optical depth}
\label{sec-tau}

The lens model is chosen to be the SIS. 
The elementary cross section for lensing is \cite{Sereno:2011ty}
\begin{equation}
    \label{eq-sis-sc}
    s_{cr}=16\pi^3\sigma^4\left( \frac{D_\text{L}D_\text{LS}}{D_\text{S}} \right)^2(y_\text{max}^2-y_\text{min}^2).
\end{equation}
Here, $y=\beta/\theta_\text{E}$, and $y_\text{max}$ is its maximal value, determined by requirement that lensed GW signals could be detected as displaying  the beat pattern. Concerning the minimal value, arising when the geometric optics approximation breaks down, we assume $y_\text{min}=0$ as suggested by Ref.~\cite{Sereno:2011ty}. 
To determine $y_\text{max}$, one first notes that $y_\text{max}\le1$ \cite{gravlens1992}. 
Second, the observed SNR of the GW signal $\rho$ should be greater than a threshold usually assumed as $\rho_\text{th}=8$. 
This might be too low to extract useful information from  the beat pattern. However, for our purpose it would be sufficient to assume this standard value. It can be easily adjusted in the following calculations. 
Third, the time delay $\Delta t=y\Delta t_z$, with \cite{Sereno:2010dr}
\begin{equation}
    \label{eq-dtz}
    \Delta t_z\equiv32\pi^2\sigma^4\frac{D_\text{L}D_\text{LS}}{D_\text{S}}(1+z_\text{L}),
\end{equation}
should be small enough, say less than $\Delta t_m=1$ month.
Now, according to Eqs.~\eqref{eq-hf} and \eqref{eq-def-snr}, one knows that 
\begin{equation}
    \label{eq-cal-snr}
        \rho_\text{obs}^2=4\int_0^\infty\frac{\mu_+^2+\mu_-^2+2\mu_+\mu_-\cos(2\pi f\Delta t)}{S_h(f)}|\tilde h_u(f)|^2\ud f.
\end{equation}
The cosine term in the integrand is highly oscillating compared to $|\tilde h_u(f)|$ in the frequency domain, so one may ignore it for the purpose of estimating the lensing rate, and the total GW SNR is $\rho_\text{obs}\approx\sqrt{\mu_+^2+\mu_-^2}\rho_\text{int}=\sqrt{2/y}\rho_\text{int}$ with $\rho_\text{int}$ the intrinsic SNR, 
\begin{equation}
    \label{eq-def-int-snr}
    \rho_\text{int}^2=4\int_0^\infty \frac{|\tilde h_u(f)|^2}{S_h(f)}\ud f.
\end{equation}
Therefore, one has
\begin{equation}
    \label{eq-ymax}
    y_\text{max}=\text{min}\left\{y_1,\frac{\Delta t_m}{\Delta t_z}\right\},
\end{equation}
with $y_1=\text{min}\{1,2\rho^2_\text{int}/\rho^2_\text{th}\}$.

Since (B-)DECIGO will operate for a limited amount of time, the actual cross section used to calculate the optical depth is \cite{Sereno:2011ty}
\begin{equation}
    s^*_{cr}=16\pi^3\sigma^4\left( \frac{D_\text{L}D_\text{LS}}{D_\text{S}} \right)^2\left(y_\text{max}^2-\frac{2\Delta t_z}{3T_s}y_\text{max}^3\right),
\end{equation}
where $T_s$ is the survey time, and set to 4 years \cite{Piorkowska-Kurpas:2020mst}.
The differential optical depth is given by \cite{Sereno:2010dr}
\begin{equation}
    \label{eq-dtau}
    \frac{\pd^2\tau}{\pd z_\text{L}\pd\sigma}=\frac{\ud n}{\ud\sigma}s^*_{cr}\frac{\ud t}{\ud z_\text{L}},
\end{equation}
where $\tau$ is the optical depth, $z_{\text{L}}$ is the redshift of the lens, $n$ is the lens number density, $t$ is the cosmological time and 
$\ud n/\ud\sigma$ is modeled  as a modified Schechter function \cite{Choi:2006qg}
\begin{equation}
    \label{eq-msf}
    \frac{\ud n}{\ud\sigma}=\frac{n_*}{\sigma_*}\frac{\beta}{\Gamma(\alpha/\beta)}\left( \frac{\sigma}{\sigma_*} \right)^{\alpha-1}\exp\left[ -\left( \frac{\sigma}{\sigma_*} \right)^\beta \right]
\end{equation}
where $\Gamma(x)$ is the gamma function, and 
\begin{gather}
    n_*=8.0\times10^{-3}h^3\text{ Mpc}^{-3},\quad\sigma_*=161\pm5\text{ km/s},\\ 
    \alpha=2.32\pm0.10,\quad\beta=2.67\pm0.07.
\end{gather}
As one can check, more than 99.8\% galaxies have $\sigma>10$ km/s. 
With this, one can estimate the curvature radius of a galaxy, i.e., on the order of $10^{9}$ m, assuming $\sigma=10\text{ km/s},\, z_\text{L}=1$, and $z_\text{S}=2$ \footnote{We choose these values because the probabilities for lensed GWs from the neutron star-neutron star mergers and the black hole-black hole mergers peak at redshifts about 2 and 4, respectively \cite{Yang:2019jhw}.}.
This curvature radius is much greater than the GW wavelength at around $0.1$ Hz, which is roughly the most sensitive frequency for (B-)DECIGO.
Indeed, the geometric optics is a good approximation.

By the definition of the redshift \cite{Weinberg:2008zzc},
\begin{equation}
    \label{eq-zdef}
    1+z=\frac{a_0}{a},
\end{equation}
one can determine the final factor in Eq.~\eqref{eq-dtau}, which is
\begin{equation}
    \label{eq-dtdz}
    \frac{\ud t}{\ud z_\text{L}}=-\frac{1}{(1+z_\text{L})H_\text{L}},\quad H_\text{L}=H(z_\text{L}).
\end{equation}
So now, one can calculate the differential optical depth $\ud\tau/\ud z_\text{L}$ using Eq.~\eqref{eq-dtau},
\begin{equation}
    \label{eq-dif-tau-1}
    \frac{\ud\tau}{\ud z_\text{L}}=\int_0^\infty\frac{\pd^2\tau}{\pd z_\text{L}\pd \sigma}\ud\sigma.
\end{equation}
The complexity of Eqs.~\eqref{eq-ymax} and \eqref{eq-dtau} makes the above integration very difficult. 
But since $\Delta t_z$ is an increasing function of $\sigma$ according to \eqref{eq-dtz}, there exists a value $\sigma_1$ such that if $\sigma<\sigma_1$, $y_\text{max}=y_1$, while if $\sigma\ge\sigma_1$, $y_\text{max}=\Delta t_m/\Delta t_z$.
This $\sigma_1$ is given by 
\begin{equation}
    \sigma_1=\left[\frac{\Delta t_m}{32\pi^2y_1}\frac{D_\text{S}}{D_\text{L}D_\text{LS}}\frac{1}{1+z_\text{L}}\right]^{1/4},
\end{equation}
obtained from the condition $y_1=\Delta t_m/\Delta t_z$.
Then, one can carry out the integration \eqref{eq-dif-tau-1} by dividing  the integration range into two parts, separated by $\sigma_1$. 
This gives
\begin{widetext}
 \begin{equation}
    \label{eq-dif-tau}
    \begin{split}
    \frac{\ud \tau}{\ud z_\text{L}}=&\frac{16\pi^3y_1^2n_*\sigma_*^4}{\Gamma(\alpha/\beta)}\frac{(1+z_\text{L})^2}{H_\text{L}}\left( \frac{D_\text{L}D_\text{LS}}{D_\text{S}} \right)^2
    \left[ \Gamma\left( \frac{\alpha+4}{\beta} \right)\mathcal P\left( u_1,\frac{\alpha+4}{\beta} \right) -\frac{2\Delta t_*y_1}{3T_s}\times\right.\\
    &\left.\Gamma\left( \frac{\alpha+8}{\beta} \right)\mathcal P\left( u_1,\frac{\alpha+8}{\beta} \right)\right]+\frac{(\Delta t_m)^2n_*}{64\pi\sigma_*^4\Gamma(\alpha/\beta)}
    \frac{1}{H_\text{L}}\left( 1-\frac{2\Delta t_m}{3T_s} \right)\mathcal Q\left( u_1,\frac{\alpha-4}{\beta} \right),
    \end{split}
\end{equation}   
\end{widetext}
where $\Delta t_*$ is given by Eq.~\eqref{eq-dtz} with $\sigma$ replaced by $\sigma_*$, $u_1=(\sigma_1/\sigma_*)^\beta$, and 
\begin{gather}
    \mathcal P(x,a)=\frac{1}{\Gamma(a)}\int_0^x\xi^{a-1}e^{-\xi}\ud\xi,\\ 
    \mathcal Q(x,a)=\int_x^\infty\xi^{a-1}e^{-\xi}\ud\xi,
\end{gather}
are the incomplete gamma functions.
The optical depth is thus 
\begin{equation}
    \label{eq-tau}
    \tau(z_\text{S})=\int_0^{z_\text{S}}\frac{\ud\tau}{\ud z_\text{L}}\ud z_\text{L}.
\end{equation}
This integration can be performed numerically.
Note that $\tau$ actually depneds on $\rho_\text{th}$.


\section{Lensing rates}
\label{sec-br}

In this section, we will estimate the yearly detection rates of lensed GW events displaying the beat pattern by generalizing the method given in Ref.~\cite{Ding:2015uha}. 
So we first present the method, and then display the new results.

We consider GWs emitted by the double compact objects (DCOs), which include NS-NS, black hole-neutron star binaries (BH-NS) and  BH-BH.
Following Ref.~\cite{Ding:2015uha}, we use the locally measured intrinsic coalescence rate $\dot n_0(z)$ for DCOs at the redshift $z$, discussed in Ref.~\cite{Dominik:2013tma} and the data (more specifically, the so-called ``rest frame rates'' in cosmological scenario) from the website \href{https://www.syntheticuniverse.org/}{https://www.syntheticuniverse.org/}.
The intrinsic rate $\dot n_0(z)$ was calculated based on well-motivated assumptions about star formation rate, galaxy mass distribution, stellar populations, their metallicities and galaxy metallicity evolution with redshift (``low-end'' and ``high-end'' cases). 
The binary system evolves from zero-age main sequence to the compact binary formation after supernova (SN) explosions.
Since the formation of the compact object is related to the physics of common envelope (CE) phase of evolution and on the SN explosion mechanism, and both of them are uncertain to some extent, Ref.~\cite{Dominik:2013tma} considered four scenarios: standard one -- based on conservative assumptions, and three of its modifications --- Optimistic Common Envelope (OCE), delayed SN explosion and high BH kicks scenario. 
For more details, see \cite{Dominik:2013tma} and references therein. 
The chirp masses ($\mathcal M_0$) are assumed: $1.2M_\odot$ for NS-NS, $3.2M_\odot$ for BH-NS, and $6.7M_\odot$ for BH-BH.
These values are the average chirp masses for these different binary systems given by the population synthesis \cite{Dominik:2012kk}. 
However, these were values obtained under the assumption of solar metallicity of initial binary systems. Such scenario was absolutely right guess before the first detection of GWs. Now, the data gathered by the LIGO/Virgo detectors have  significantly modified these guesses demonstrating that observed chirp masses (particular of BH-BH systems) are much higher. Therefore, guided by the real data collected so far we will adopt different values. 
According to \cite{LIGOScientific:2018mvr}, we will assume the median  value of BH-BH systems chirp masses reported in their Table III. Since the data on BH-NS systems is more scarce, we will take the value of \cite{LIGOScientific:2020stg}. 
In summary,  we take the following values as representatives for typical chirp masses $\mathcal M_0$ of DCO inspiralling systems:
 $1.2\;M_\odot$ for NS-NS, $6.09\; M_\odot$ for BH-NS, and
$24.5\;M_\odot$ for BH-BH.  

Since a few dozens of GW events have been observed, LIGO/Virgo collaboration inferred the merger rates \cite{LIGOScientific:2018mvr,Abbott:2020gyp}. 
We will also present the lensing rates using the inferred merger rates. 
Note that the merger rates for NS-NS and BH-BH binaries still suffer from large error bars, and there is only an upper bound on the BH-NS merger rate. 
In addition, the most distant source is at $z_\text{S}=0.8$.
Therefore, we relegate the results in Appendix~\ref{app-inf}.
In the following, we will continue to use the merger rate $\dot n_0$ reported in Ref.~\cite{Dominik:2013tma}.

Matched filtering is used to identify GW events. 
The \emph{intrinsic} SNR for a single detector can be approximately determined with  \cite{Taylor:2012db}
\begin{equation}
    \label{eq-snr}
    \rho=8\Theta\frac{r_0}{d_\text{L}(z_\text{S})}\left( \frac{\mathcal M_z}{1.2M_\odot} \right)^{5/6}\sqrt{\zeta(f_\text{max})},
\end{equation}
where $\Theta$ is the orientation factor, $\mathcal M_z = (1+z_\text{S}) \mathcal M_0$ is the chirp mass registered by the detector,  $d_\text{L}(z_\text{S})$ is the luminosity distance, and finally, $r_0$ is the detector's characteristic distance. 
The function $\zeta(f_\text{max})$ is 
\begin{equation}
    \zeta(f_\text{max})=\frac{1}{x_{7/3}}\int_0^{2f_\text{max}}\frac{(\pi M_\odot)^2}{(\pi M_\odot f)^{7/3}S_h(f)}\ud f,
\end{equation}
where $x_{7/3}$ is nothing but the above integration with the upper limit being infinity.
Since DCO inspiralling systems studied in this work pass the sensitivity bands of (B-)DECIGO, one assumes that $\zeta(f_\text{max})=1$ \cite{Piorkowska-Kurpas:2020mst}.
The characteristic distance parameter $r_0$ is determined by 
\begin{equation}
    \label{eq-def-r0}
    r_0^2=\frac{5}{192\pi^{4/3}}\left( \frac{3GM_\odot}{20} \right)^{5/3}\int_0^\infty\frac{\ud f}{f^{7/3}S_h(f)}.
\end{equation}
It depends only on the noise spectrum $S_h(f)$ of the detector, so it also characterizes the detector's sensitivity.
The larger it is, the more sensitive the detector is.
By the above equation, one finds out that $r_0=6709$ Mpc for DECIGO, and $r_0=535$ Mpc for B-DECIGO.

The orientation factor $\Theta$ in Eq.~\eqref{eq-snr} is defined as 
\begin{equation}
    \Theta=2\sqrt{F_+^2(1+\cos^2\iota)^2+4F_\times^2\cos^2\iota},
\end{equation}
where $F_+$ and $F_\times$ are the antenna pattern functions for the $+$ and $\times$ polarizations, respectively, given by \cite{Poisson2014}
\begin{gather}
    F_+=\frac{1+\cos^2\theta}{2}\cos2\phi\cos2\psi-\cos\theta\sin2\phi\sin2\psi,\\ 
    F_\times=\frac{1+\cos^2\theta}{2}\cos2\phi\sin2\psi+\cos\theta\sin2\phi\cos2\psi.
\end{gather}
In these expressions, $\theta$ and $\phi$ are the polar and azimuthal angles of the spherical coordinate system, which centers at the detector and whose $z$-axis is perpendicular to the detector plane. 
$\iota$ is the inclination angle between the GW propagation direction and the angular momentum of the binary system, and finally, $\psi$ is called the polarization angle.
With a single interferometer, one cannot measure $\Theta$, but one can infer its probability distribution $P(\Theta)$.
Since averaged over a lot of binaries, $\cos\theta,\phi/\pi,\cos\iota,$ and $\psi/\pi$ are uncorrelated and uniformly distributed over the range $(-1,1)$, $P(\Theta)$  is approximated by \cite{Finn:1995ah}
\begin{equation}
    P(\Theta)=\frac{5\Theta(4-\Theta)^3}{256},
\end{equation}
for $0<\Theta<4$, and $P(\Theta)=0$, otherwise.
It is easy to verify that $P(\Theta)$ peaks at $\Theta=1$.

Now, it is possible to understand the physical meaning of $r_0$ by first substituting Eq.~\eqref{eq-def-r0} into Eq.~\eqref{eq-snr}, and then solving for $d_\text{L}^2$,
\begin{equation}
    \label{eq-dl2}
    d_\text{L}^2=\frac{64\Theta^2}{\rho^2}\frac{5}{192\pi^{4/3}}\left( \frac{G\mathcal M_z}{8} \right)^{\frac{5}{3}}\int_0^{2f_\text{max}}\frac{\ud f}{f^{7/3}S_h(f)}.
\end{equation}
One immediately finds out that $d_\text{L}$ becomes $r_0$ once one sets $\rho=8$ (the standard threshold), $\Theta=1$, $\mathcal M_z=1.2M_\odot$, and $f_\text{max}=+\infty$.
This means that $r_0$ is the luminosity distance of a ``fiducial'' source whose redshifted chirp mass is $1.2M_\odot$, which is at the very orientation such that $\Theta=1$ (i.e., the most probable orientation), which, hypothetically, emits GW only at the quadruple order all the time ($f_\text{max}=+\infty$), and whose GW signal has the SNR of 8.
It is easy to understand that a more sensitive interferometer can detect a more distant fiducial source. 
Therefore, $r_0$ characterizes the sensitivity of a detector.

The differential beat rate is given by \cite{Ding:2015uha}
\begin{equation}
    \label{eq-dif-rate}
\frac{\pd^2\dot N}{\pd z_\text{S}\pd\rho}=\frac{4\pi\dot n_0(z_\text{S})}{(1+z_\text{S})^3}\frac{d_\text{L}^2(z_\text{S})}{H(z_\text{S})}\tau(z_\text{S})P(x(z_\text{S},\rho))\frac{x(z_\text{S},\rho)}{\rho},
\end{equation}
where the intrinsic coalescence rate $\dot n_0$ has been introduced at the beginning of this section, and
\begin{equation}
    \label{eq-def-x}
    x(z,\rho)=\frac{\rho}{8}(1+z)^{-5/6}\frac{d_\text{L}(z)}{r_0}\left( \frac{1.2M_\odot}{\mathcal M_0} \right)^{5/6}.
\end{equation}
The yearly detection rate is thus
\begin{equation}
    \label{eq-rate-tot}
    \dot N=\int_0^{z_\text{max}}\ud z_\text{S}\int_0^\infty\ud\rho\frac{\pd^2\dot N}{\pd z_\text{S}\pd\rho},
\end{equation}
and the differential yearly detection rates are defined to be 
\begin{gather}
    \frac{\pd\dot N}{\pd\rho}=\int_0^{z_\text{max}}\ud z_\text{S}\frac{\pd^2\dot N}{\pd z_\text{S}\pd\rho},\\ 
    \frac{\pd\dot N}{\pd z_\text{S}}=\int_0^\infty\ud\rho\frac{\pd^2\dot N}{\pd z_\text{S}\pd\rho}.
\end{gather}
With the method presented above, one can obtain the following new results.

\begin{figure}[h]
    \centering
    \includegraphics[width=0.45\textwidth]{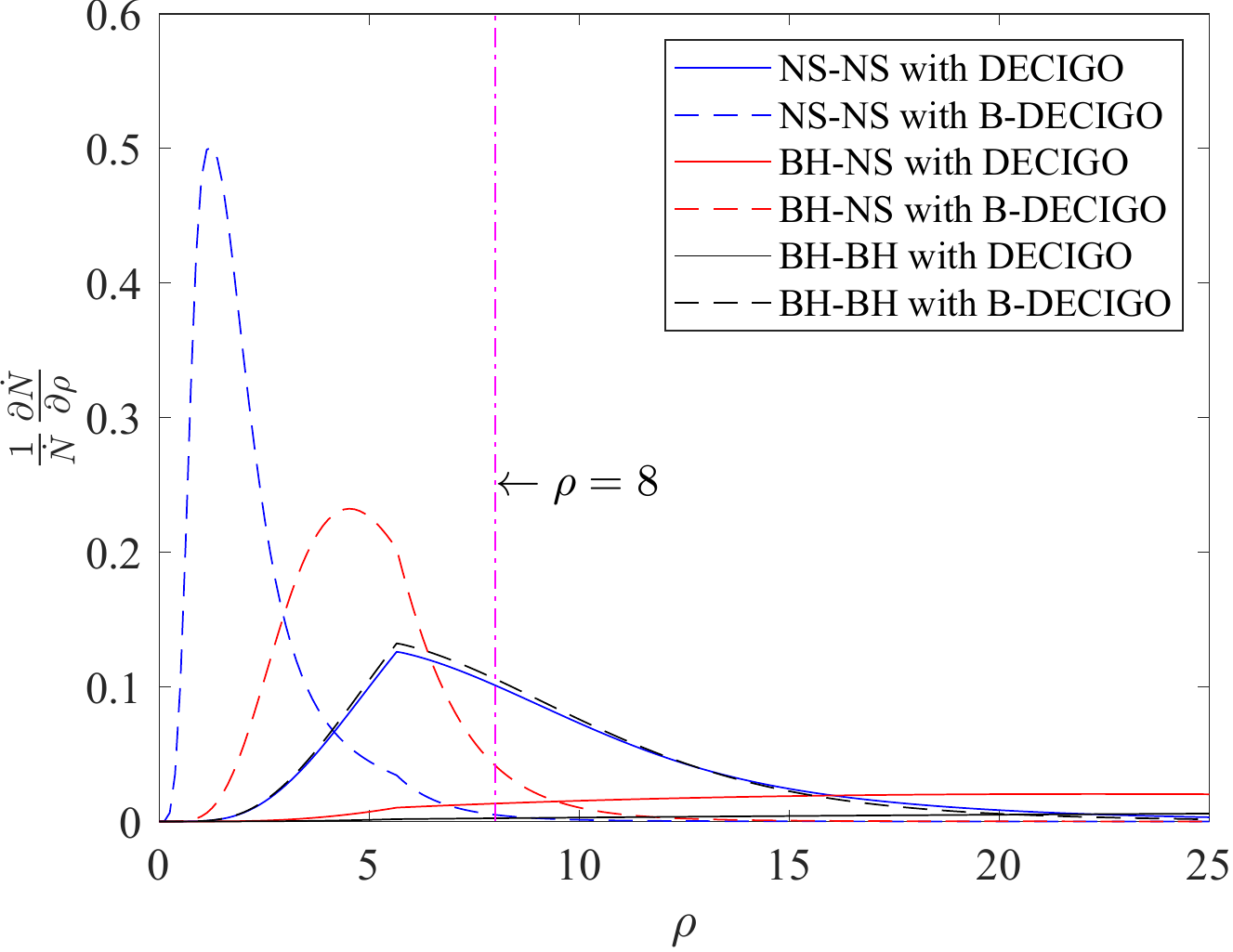}
    \caption{The relative differential detection rates $\frac{1}{\dot N}\frac{\pd\dot N}{\pd\rho}$ v.s. $\rho$ for (B-)DECIGO.
    The evolutionary model for DCOs is the standard one with the ``low-end'' metallicity scenario.
    The purple dot-dashed line is at $\rho=8$.}
    \label{fig-reldrrho}
\end{figure}

Figure~\ref{fig-reldrrho} shows the relative differential detection rates $\frac{1}{\dot N}\frac{\pd\dot N}{\pd\rho}$ v.s. $\rho$ for (B-)DECIGO.
The evolutionary model for DCOs is the standard one with the ``low-end'' metallicity scenario.
As one can see low SNR events of three different types of DCOs dominate for B-DECIGO.
For DECIGO, although the relative differential rate for NS-NS type DCOs peaks at a small SNR ($\rho<\rho_\text{th}$), the curves for the remaining types of DOCs are more flat and reach maximum at higher SNRs. 

\begin{figure}[h]
    \centering
    \includegraphics[width=0.45\textwidth]{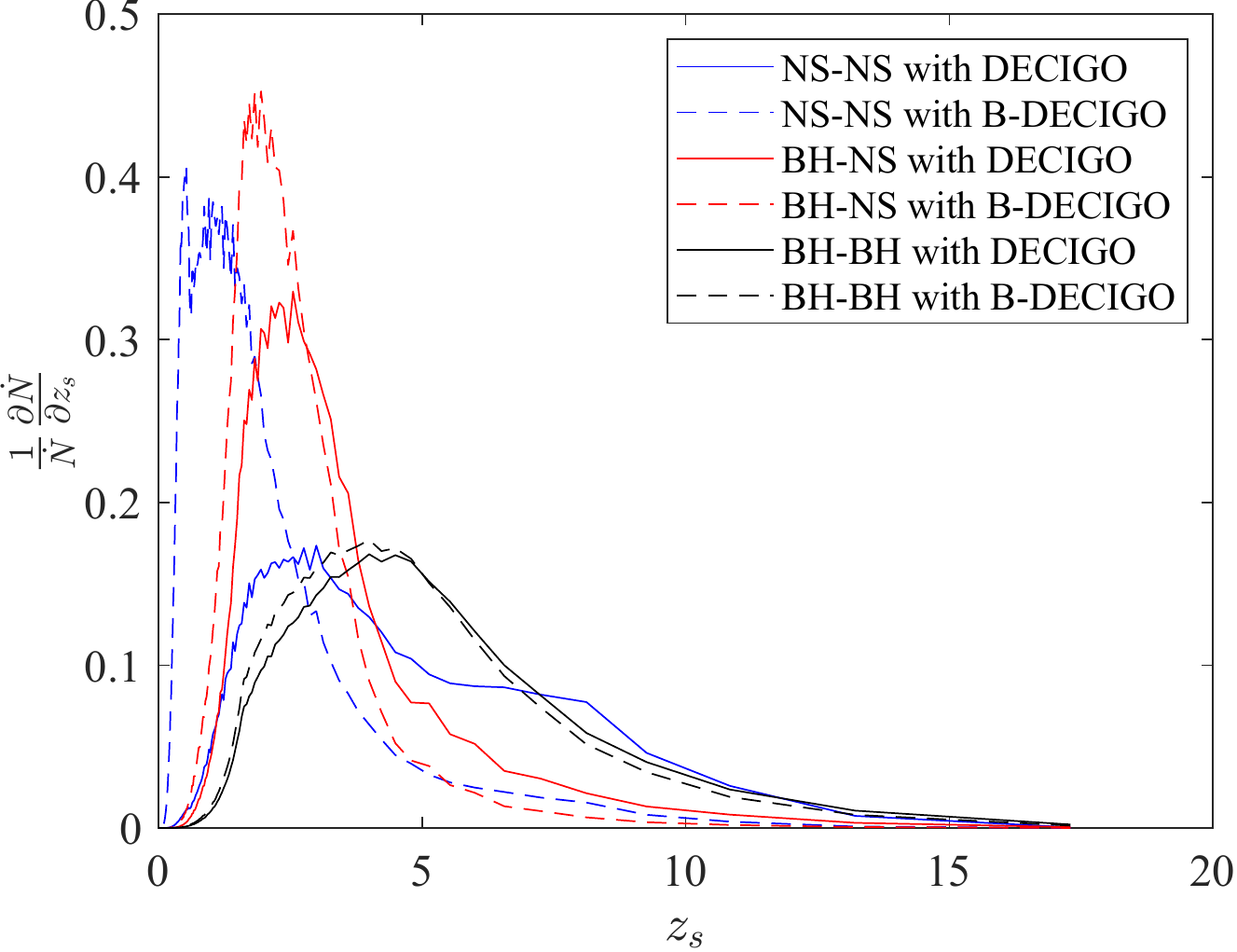}
    \caption{The relative differential detection rates $\frac{1}{\dot N}\frac{\pd\dot N}{\pd z_\text{S}}$ v.s. $z_\text{S}$ for (B-)DECIGO.
    The evolutionary model for DCOs is the standard one with the ``low-end'' metallicity scenario.}
    \label{fig-reldr}
\end{figure}

Figure~\ref{fig-reldr} displays the relative differential detection rates $\frac{1}{\dot N}\frac{\pd\dot N}{\pd z_\text{S}}$ v.s. $z_\text{S}$ for (B-)DECIGO.
The evolutionary model for DCOs is also the standard one with the ``low-end'' metallicity scenario.
From this figure, one can see that lensed GW events observable in DECIGO are dominated by NS-NS and BH-NS binaries at $z_\text{S}=2\sim 4$ and BH-BH binaries at $z_\text{S}=4\sim5$. 
On the other hand, differential lensing rates for B-DECIGO peak at the slightly lower redshifts, respectively.
The earlier launch of B-DECIGO would  provide valuable information. 

Table~\ref{tab-rates-dec} displays the yearly detection rate $\dot N$ for lensed GW events with the beat pattern from the inspiraling DCOs of different classes. 
As shown in the table, we consider all four scenarios with 
both the low-end and high-end metallicity evolutions assumed. 
\begin{table}[h]
\begin{center}  
\begin{tabular}{c|c|c|c|c}
\hline 
$\dot N$ & Stand. & Opt. CE & Del. SN &  BH kicks \\
\hline 
 NS-NS &&&& \\
low-end metallicity & $10.2$ & $83.8$ & $10.7$ & $10.3$ \\
high-end metallicity & $10.3$ & $88.5$ & $11.0$ & $10.6$ \\
\hline 
 BH-NS &&&& \\
low-end metallicity & $5.9$ & $9.5$ & $2.9$ & $0.7$ \\
high-end metallicity & $5.1$ & $9.5$ & $2.5$ & $0.6$ \\
\hline
 BH-BH &&&& \\
low-end metallicity & $111.5$ & $275.4$ & $93.8$ & $8.0$ \\
high-end metallicity & $92.0$ & $255.9$ & $76.9$ & $6.4$ \\
\hline
Total &&&& \\
low-end metallicity & $127.6$ & $368.7$ & $107.4$ & $19.0$ \\
high-end metallicity & $107.4$ & $353.9$ & $90.4$ & $17.6$ \\
\hline
\end{tabular}
\caption{Yearly detection rates for lensed GW events exhibiting the beat pattern from inspiralling DCOs of different classes under different evolutionary scenarios, assuming ``low-end'' and ``high-end'' metallicity evolutions.
Predictions for DECIGO. }
\label{tab-rates-dec}
\end{center}
\end{table}
From this table, one finds out that lensed GWs generated by BH-BH binary systems dominate in most cases, except for the High BH kicks scenario, for which lensed GWs from NS-NS binaries contribute the most.
One interesting result is that in all cases, there are at least 10 lensed GW events with the beat pattern from the NS-NS binaries per year.
This creates possibility that at least for some of them, electromagnetic counterparts could be detected, allowing to identify the host galaxy and measure the redshift. Hence, the cosmological parameters could be measured from them according to Ref.~\cite{Hou:2019dcm}.  
Of course, one should also try to take advantage of the dominating BH-BH binary systems using statistical methods as discussed in Ref.~\cite{Sereno:2011ty}.

Table~\ref{tab-rates-bdec} shows the yearly detection rate for lensed GW events with the beat pattern for B-DECIGO.
Compared with Table~\ref{tab-rates-dec}, it has similar features but the rates are smaller. 
This is due to the lower designed sensitivity. 
However, there is still a considerable amount of lensed events dominated by signals from BH-BH systems. The only exception is the High BH kicks scenario,
leading to suppression of BH-BH formation rate. In such a case the perspectives for the B-DECIGO to detect lensed GW with a beat pattern are poor. 

\begin{table}[h]
\begin{center}  
\begin{tabular}{c|c|c|c|c}
\hline 
$\dot N$ & Stand. & Opt. CE & Del. SN &  BH kicks \\
\hline 
NS-NS &&&& \\
low-end metallicity & $0.05$ & $0.63$ & $0.06$ & $0.06$ \\
high-end metallicity & $0.09$ & $0.59$ & $0.10$ & $0.09$ \\
\hline 
BH-NS &&&& \\
low-end metallicity & $1.61$ & $2.98$ & $0.83$ & $0.17$ \\
high-end metallicity & $1.19$ & $2.67$ & $0.62$ & $0.13$ \\
\hline
BH-BH &&&& \\
low-end metallicity & $82.8$ & $216.7$ & $69.0$ & $5.65$ \\
high-end metallicity & $66.6$ & $197.7$ & $55.14$ & $4.43$ \\
\hline
Total &&&& \\
low-end metallicity & $84.46$ & $220.3$ & $69.89$ & $5.88$ \\
high-end metallicity & $67.88$ & $201.0$ & $55.86$ & $4.65$ \\
\hline
\end{tabular}
\caption{Yearly detection rates for lensed GW events with the beat pattern from inspiralling DCOs of different classes under different evolutionary scenarios, assuming ``low-end'' and ``high-end'' metallicity evolutions.
Predictions for B-DECIGO. }
\label{tab-rates-bdec}
\end{center}
\end{table}

One may concern that in order to detect the beat patterns with enough accuracy to determine the luminosity distance, the lens mass and cosmological parameters, SNR threshold $\rho_\text{th}$ should be bigger than usually assumed value $\rho_\text{th}=8.$ 
So we also calculate the detection rates by increasing $\rho_\text{th}$ by a factor of 10 or even 100, which are listed in Appendix~\ref{app-dr}.
From there, one knows that even with $\rho_\text{th}=80$, DECIGO could still detect a lot of lensed events with beat patterns, still dominated by BH-BH events.
Unfortunately, rates for NS-NS events drop quite a lot.
The detection ability of B-DECIGO decreases substantially, and it is probably not feasible to use it to detect lensing events.
At the even higher threshold SNR $\rho_\text{th}=800$, all lensing rates are diminishingly small.

\section{Conclusion}
\label{sec-con}

In this work, we analyze how many lensed GW events with the beat pattern can be detected by (B-)DECIGO every year. 
It turns out that there are many more lensed events from DCOs than those observable by LISA (with $\rho_\text{th}=8$).
Among different binary types of DCOs, BH-BH contribution is dominating  in most evolutionary models of DCOs.
Nevertheless, there is still a considerable number of lensed GWs expected from NS-NS and BH-NS binaries, which can be used together with their electromagnetic counterparts to study the cosmology accurately. 
In fact, the lensed GWs from BH-BH binaries are also valuable with the statistics method \cite{Sereno:2011ty}, even though there are no electromagnetic counterparts.
At $\rho_\text{th}=80$, the detection ability of DECIGO decreases a little, while that of B-DECIGO drops dramatically.
Of  course, at the even higher $\rho_\text{th}=800$, neither of them is suitable for detecting beat patterns.

Another point worth mentioning is that it is very advantageous  to study cosmology with the beat pattern, because high redshift binaries ($z_\text{S}=3\sim 6$) contribute a lot to the total detection rates. 
In Ref.~\cite{Hou:2019dcm}, the authors discussed how to use the beat pattern to measure the luminosity distance of the GW source, the mass of the lens, and some cosmological parameters (e.g., $H_0$).
In principle, these measurements can be very accurate.
However, these studies were based on the simple lens models: the point-mass model and SIS. One may expect that similar opportunity will emerge in more complicated and realistic lens mass profiles. This deserves further study. 
Moreover some complications arising in realistic situation were also omitted, such as the small SNRs for some GW events, the intrinsic scatter in the lens profile and the cosmic shear, etc..
One has to properly address these issues in the real measurements in order to guarantee accuracy of the method.  
In this work, we only estimate the lensing rate, which at the order of magnitude level would not be affected much by these factors. 
Our predictions raise hopes of detecting beat patterns in forthcoming (B-)DECIGO missions and motivates to undertake more realistic studies of this phenomenon.

\begin{acknowledgements}
  This work was supported by the National Natural Science Foundation of China under Grants No.~11633001, No.~11673008, No.~11922303, and No.~11920101003 and the Strategic Priority Research Program of the Chinese Academy of Sciences, Grant No. XDB23000000.
  SH was supported by Project funded by China Postdoctoral Science Foundation (No.~2020M672400).
  HY was supported by Initiative Post-docs Supporting Program (No. BX20190206) and Project funded by China Postdoctoral Science Foundation (No.~2019M660085). 
MB was supported by the Key Foreign Expert Program for the Central Universities No. X2018002. 
\end{acknowledgements}

\appendix

\section{Lensing rates at higher threshold SNRs}
\label{app-dr}

Although it is a standard practice to assume $\rho_\text{th}=8$, we want to increase it in order to make sure that it is easier to obtain fairly accurate information from the beat pattern to make the measurements proposed in Ref.~\cite{Hou:2019dcm}.
In this section, we present the detection rates at higher threshold SNRs, i.e., $\rho_\text{th}=80$ and $800$.
Table \ref{tab-rates-dec-80} displays the lensing rates for DECIGO when $\rho_\text{th}=80$. 
As one expects, the rates decrease, compared with Table \ref{tab-rates-dec}. 
\begin{table}[h]
\begin{center}  
\begin{tabular}{c|c|c|c|c}
\hline 
$\dot N$ & Stand. & Opt. CE & Del. SN &  BH kicks \\
\hline 
 NS-NS &&&& \\
low-end & $0.12$ & $1.41$ & $0.13$ & $0.12$ \\
high-end & $0.19$ & $1.31$ & $0.22$ & $0.20$ \\
\hline 
 BH-NS &&&& \\
low-end & $2.31$ & $4.15$ & $1.18$ & $0.25$ \\
high-end & $1.78$ & $3.84$ & $0.93$ & $0.20$ \\
\hline
 BH-BH &&&& \\
low-end & $91.3$ & $234.4$ & $76.3$ & $6.33$ \\
high-end & $74.0$ & $215.2$ & $61.5$ & $5.00$ \\
\hline
Total &&&& \\
low-end & $93.7$ & $240.0$ & $77.6$ & $6.70$ \\
high-end & $76.0$ & $220.4$ & $62.6$ & $5.40$ \\
\hline
\end{tabular}
\caption{Yearly detection rates for lensed GW events exhibiting the beat pattern from inspiralling DCOs of different classes under different evolutionary scenarios, assuming ``low-end'' and ``high-end'' metallicity evolutions, at $\rho_\text{th}=80$.
Predictions for DECIGO. }
\label{tab-rates-dec-80}
\end{center}
\end{table}
Table \ref{tab-rates-bdec-80} contains the lensing rates calculated for B-DECIGO at $\rho_\text{th}=80$. 
Very interestingly, they are much smaller than the corresponding ones by 2 to 4 orders of magnitude in Table \ref{tab-rates-bdec}.
\begin{table}[h]
\begin{center}  
\begin{tabular}{c|c|c|c|c}
\hline 
$\dot N$ & Stand. & Opt. CE & Del. SN &  BH kicks \\
\hline 
NS-NS &&&& \\
low-end & $6.8\times10^{-6}$ & $7.2\times10^{-5}$ & $7.7\times10^{-6}$ & $7.0\times10^{-6}$ \\
high-end & $1.2\times10^{-5}$ & $6.8\times10^{-5}$ & $1.3\times10^{-5}$ & $1.2\times10^{-5}$ \\
\hline 
BH-NS &&&& \\
low-end & $8.1\times10^{-4}$ & $0.002$ & $4.2\times10^{-4}$ & $8.0\times10^{-5}$ \\
high-end & $5.0\times10^{-4}$ & $0.002$ & $2.7\times10^{-4}$ & $6.0\times10^{-5}$ \\
\hline
BH-BH &&&& \\
low-end & $0.62$ & $2.7$ & $0.49$ & $0.03$ \\
high-end & $0.41$ & $2.0$ & $0.32$ & $0.02$ \\
\hline
Total &&&& \\
low-end & $0.63$ & $2.7$ & $0.49$ & $0.03$ \\
high-end & $0.41$ & $2.0$ & $0.32$ & $0.02$ \\
\hline
\end{tabular}
\caption{Yearly detection rates for lensed GW events with the beat pattern from inspiralling DCOs of different classes under different evolutionary scenarios, assuming ``low-end'' and ``high-end'' metallicity evolutions, at $\rho_\text{th}=80$.
Predictions for B-DECIGO. }
\label{tab-rates-bdec-80}
\end{center}
\end{table}
In contrast, the numbers for BH-NS and BH-BH in Tables \ref{tab-rates-dec} and \ref{tab-rates-dec-80} are very close, and those for NS-NS differ by about 2 orders of magnitude.
If one increases the threshold SNR further to $\rho_\text{th}=800$, one finds the following lensing rates for DECIGO in Table \ref{tab-rates-dec-800}.
\begin{table}[h]
\begin{center}  
\begin{tabular}{c|c|c|c|c}
\hline 
$\dot N$ & Stand. & Opt. CE & Del. SN &  BH kicks \\
\hline 
 NS-NS &&&& \\
low-end & $1.7\times10^{-5}$ & $1.8\times10^{-4}$ & $1.9\times10^{-5}$ & $1.7\times10^{-5}$ \\
high-end & $2.8\times10^{-5}$ & $1.7\times10^{-4}$ & $3.2\times10^{-5}$ & $2.9\times10^{-5}$ \\
\hline 
 BH-NS &&&& \\
low-end & $0.002$ & $0.005$ & $0.001$ & $1.9\times10^{-4}$ \\
high-end & $0.00178$ & $0.004$ & $6.7\times10^{-4}$ & $1.5\times10^{-4}$ \\
\hline
 BH-BH &&&& \\
low-end & $1.39$ & $5.67$ & $1.08$ & $0.06$ \\
high-end & $0.92$ & $4.38$ & $0.71$ & $0.04$ \\
\hline
Total &&&& \\
low-end & $1.39$ & $5.67$ & $1.08$ & $0.06$ \\
high-end & $0.92$ & $4.38$ & $0.72$ & $0.04$ \\
\hline
\end{tabular}
\caption{Yearly detection rates for lensed GW events exhibiting the beat pattern from inspiralling DCOs of different classes under different evolutionary scenarios, assuming ``low-end'' and ``high-end'' metallicity evolutions, at $\rho_\text{th}=800$.
Predictions for DECIGO. }
\label{tab-rates-dec-800}
\end{center}
\end{table}
As one can find out, the lensing rates decrease substantially by 2-6 orders of magnitude, compared with Table \ref{tab-rates-dec}.
The rates for B-DECIGO are even smaller than those in Table \ref{tab-rates-bdec-80}, and it would be impractical to use B-DECIGO to observe the GW events with beat patterns if $\rho_\text{th}=800$.
So we do not present the rates here.

This observation might be explained by computing the ``characteristic'' SNR $\rho_c$, which can be evaluated with Eq.~\eqref{eq-snr}, assuming that a binary system is at redshift $1-2$, it has the averaged chirp mass, and it is at the optimal orientation $\Theta=1$. 
The choice of the source redshift is due to the fact that merger rates of various types of binary systems peak around the chosen range \cite{Dominik:2013tma}.
The higher $\rho_c$ is, the more easily a GW event can be detected.
The ``characteristic'' SNRs $\rho_c$ for different types of binary systems and different detectors are tabulated in Table \ref{tab-rhoc}.
\begin{table}[h]
    \centering
    \bgroup
    \def\arraystretch{1.2}
   \begin{tabular}{c|c|c|c|c}
        \hline
        \multirow{2}{*}{Binary}& \multicolumn{2}{c|}{DECIGO} & \multicolumn{2}{c}{B-DECIGO}\\
        \cline{2-5}
        & $z_\text{S}=1$ & $z_\text{S}=2$ & $z_\text{S}=1$ & $z_\text{S}=2$ \\
        \hline
        NS-NS & 14.0 & 8.35 & 1.12 & 0.67\\
        BH-NS & 54.2 & 32.3 & 4.33 & 2.58 \\
        BH-BH & 173.1 & 103.2 & 13.8 & 8.23\\
        \hline
    \end{tabular}
    \egroup
   \caption{The ``characteristic'' SNRs $\rho_c$ for DECIGO and B-DECIGO at the source redshift $z_\text{S}=1$ and 2.}
    \label{tab-rhoc}
\end{table}
It turns out that $\rho_c$'s for DECIGO are generally larger than the standard threshold SNR $\rho_\text{th}=8$, but for B-DECIGO, $\rho_c$'s are smaller for NS-NS and BH-NS binaries, and is very close to $\rho_\text{th}$ for BH-BH binaries. 
This explains why the rates for BH-BH in Table \ref{tab-rates-bdec} are very similar to the relevant rates in Table \ref{tab-rates-dec}.
This also explains why the rates for BH-NS and BH-BH in Table \ref{tab-rates-dec-80} decrease a little even when the threshold SNR increases to $\rho_\text{th}=80$: 
$\rho_c$'s are still close to this new threshold.
But for NS-NS binaries, $\rho_c$'s are much smaller, so in Table \ref{tab-rates-dec-80}, their rates drop a lot. 
When $\rho_\text{th}$ is set to 800, all of $\rho_c$'s are dwarfed by this large threshold, so even for DECIGO, the rates in Table \ref{tab-rates-dec-800} are tiny.

\section{Lensing rates with the observed merger rates}
\label{app-inf}

Since a few dozens of GW events have been observed, LIGO/Virgo collaboration inferred the merger rates. 
After the first and the second observing runs, the merger rates are found to be $\mathcal R=110-3840\text{ Gpc}^{-3}\text{yr}^{-1}$ for NS-NS and $\mathcal R=9.7-101 \text{ Gpc}^{-3}\text{yr}^{-1}$ for BH-BH, and the merger rate for BH-NS is bounded from above by $\mathcal R=610\text{ Gpc}^{-3}\text{yr}^{-1}$, all at the 90\% confidential level \cite{LIGOScientific:2018mvr}.
Recently, LIGO/Virgo collaboration published the second GW transient catalog  together with the population properties \cite{Abbott:2020niy,Abbott:2020gyp}.
They updated NS-NS and BH-BH merger rates. 
It is found out that the NS-NS merger rate is $\mathcal R=320^{+490}_{-240}\text{ Gpc}^{-3}\text{yr}^{-1}$, assuming the independence of the source redshift $z$.
The BH-BH merger rate is modeled as $\mathcal R(1+z)^\kappa$, since this rate might  increase with $z$.
In the case of $\kappa=0$, the merger rate can be $\mathcal R=23.9^{+14.9}_{-8.6}\text{ Gpc}^{-3}\text{yr}^{-1}$ for the so-called POWER LAW + PEAK, BROKEN POWER LAW and MULTI PEAK mass distributions, and $\mathcal R=33^{+22}_{-12}\text{ Gpc}^{-3}\text{yr}^{-1}$ for the TRUNCATED mass distribution. 
These results were obtained without considering GW190814. 
Including it, the rate becomes $\mathcal R=58^{+54}_{-29}\text{ Gpc}^{-3}\text{yr}^{-1}$ for the POWER LAW + PEAK mass distribution.
For the details, please refer to Ref.~\cite{Abbott:2020gyp}.
In the case of $\kappa\ne0$, $\mathcal R=19.1^{+16.2}_{-9.0}\text{ Gpc}^{-3}\text{yr}^{-1}$, and $\kappa=1.3^{+2.1}_{-2.1}$ for the POWER LAW + PEAK model and $\kappa=1.8^{+2.1}_{-2.2}$ for the BROKEN POWER LAW model.
In this section, we use the inferred merger rates to compute how many lensed GW events with beat patterns can be detected a year. 
We will use the updated NS-NS rate and the upper limit on BH-NS rate without any redshift dependence.
The updated BH-BH rates will also be used with $\kappa=0, 1.3$, and $1.8$.
Although LIGO/Virgo collaboration only detected low redshift events,
we will assume the merger rate is valid up to redshift 18, which is about the maximum redshift reported in Ref.~\cite{Dominik:2013tma}. 

Since $\mathcal R$ takes different values for different models, in the actual calculation, we substitute the reference merger rate $\dot n_0=(1+z)^\kappa$ into Eq.~\eqref{eq-dif-rate}.
The resultant lensing rate is called the ``normalized'' rate. 
The actual rate is given by multiplying the normalizedone by $\mathcal R$.

At $\rho_\text{th}=8$, the normalized lensing rates are listed in Table~\ref{tab-nor8}.
From this table, one knows that DECIGO could detect about 0.188 lensed NS-NS events per year.
But this is the normalized rate. 
The actual rate is $0.188\times 320^{+490}_{-240}=60.16^{+92.12}_{-45.12}$ per year.
\begin{table}[h]
    \centering
    \bgroup
    \def\arraystretch{1.2}
    \begin{tabular}{c|c|c|c|c|c|c}
        \hline
        \multirow{2}{*}{Binary}& \multicolumn{3}{c|}{DECIGO} & \multicolumn{3}{c}{B-DECIGO} \\
        \cline{2-7}
        & $\kappa=0$ & $\kappa=1.3$ & $\kappa=1.8$& $\kappa=0$ & $\kappa=1.3$ & $\kappa=1.8$\\
        \hline
        NS-NS & 0.188 & - & - & 0.001 & - & -  \\
        BH-NS & 0.253 & - & - & 0.043 & - & - \\
        BH-BH & 0.260 & 4.04 & 13.0 & 0.186 & 2.58 & 8.05 \\
        \hline
    \end{tabular}
    \egroup
    \caption{Normalized lensing rates at $\rho_\text{th}=8$.}
    \label{tab-nor8}
\end{table}
One can similarly obtain the actual lensing rates for the remaining binary types and models.
From this table, one can also find out that as $\kappa$ increases, the lensing rates for BH-BH binaries increase.

\begin{figure}[h]
    \centering
    \includegraphics[width=0.45\textwidth]{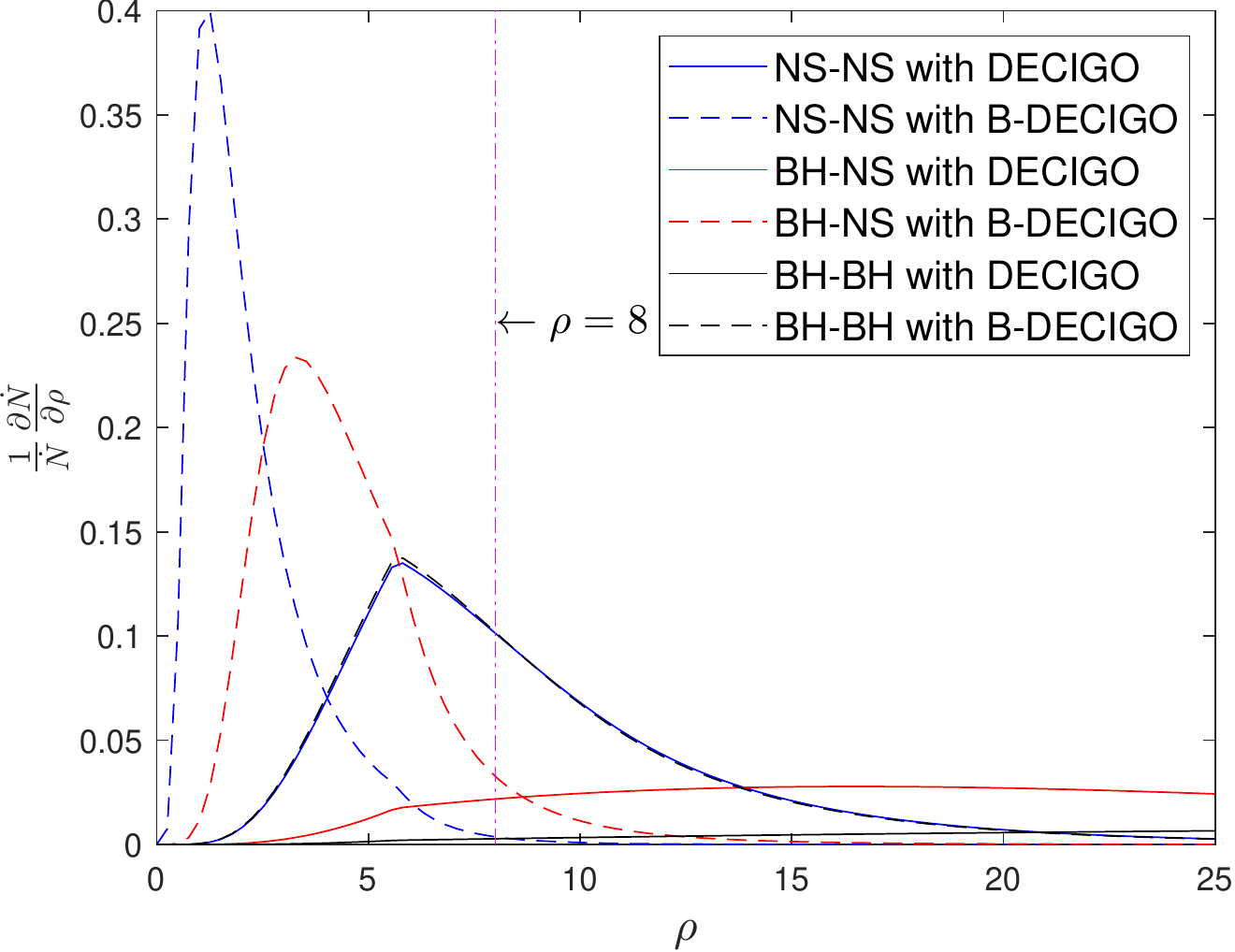}
    \includegraphics[width=0.45\textwidth]{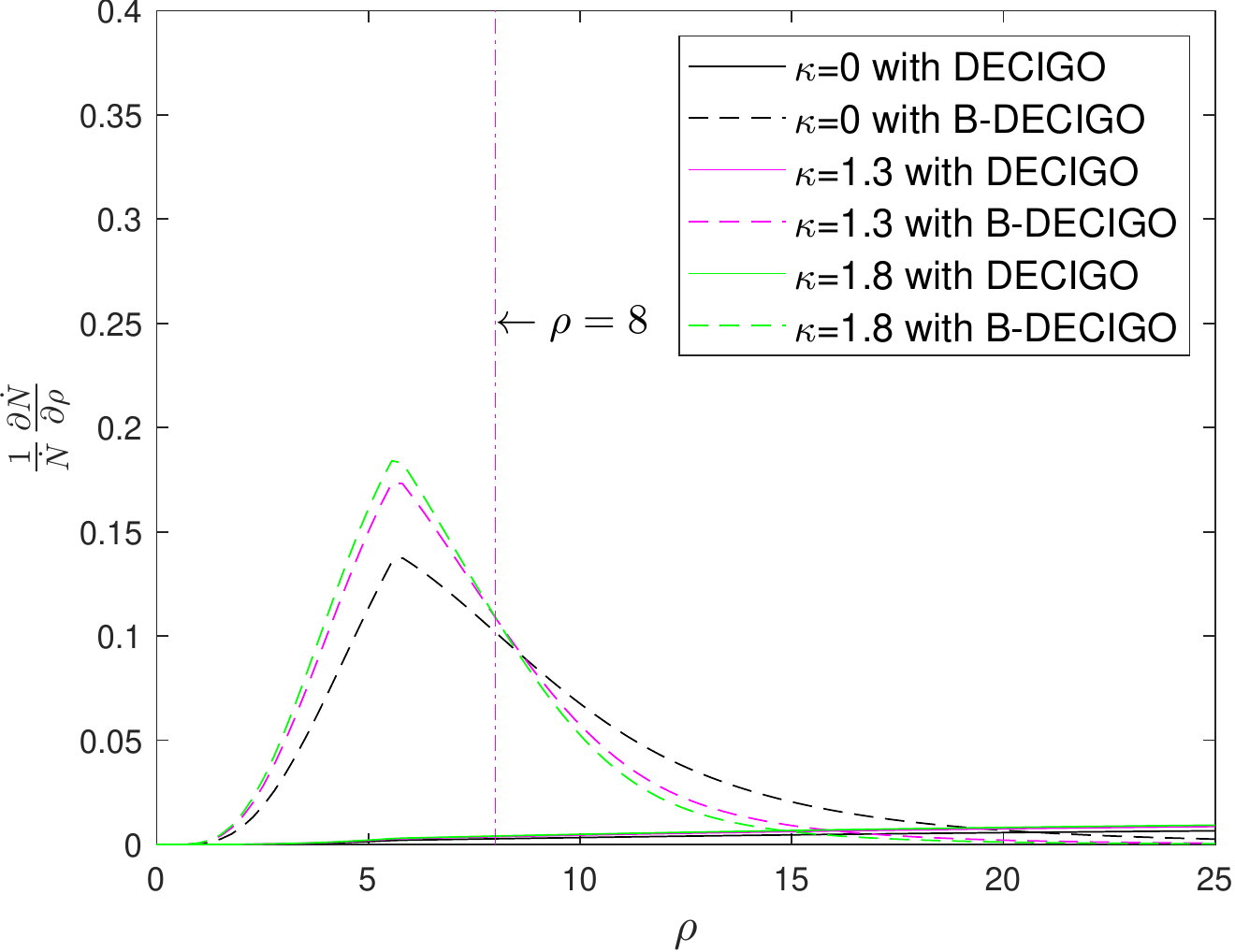}
    \caption{The relative differential detection rates $\frac{1}{\dot N}\frac{\pd\dot N}{\pd\rho}$ v.s. $\rho$ for (B-)DECIGO.
    The threshold SNR is 8.
    The upper panel shows the rates at $\kappa=0$ for various types of binary systems, and the lower panel shows the rates for BH-BH systems at various $\kappa$'s.
    The black curves in the lower panel are the same as the black ones in the upper panel.}
    \label{fig-app-rho}
\end{figure}

Figure~\ref{fig-app-rho} shows the relative differential rates $\frac{1}{\dot N}\frac{\pd\dot N}{\pd\rho}$, independent of $\mathcal R$.
In the upper panel, the rates for different types of binary systems are plotted, assuming the merger rates are independent of the redshift, i.e., $\kappa=0$.
Just like Fig.~\ref{fig-reldrrho}, B-DECIGO mainly observes events with small SNRs, while DECIGO's curves for BH-NS and BH-BH are flatter.
Of course, DECIGO is also more sensitive to NS-NS events with small SNRs.
Similar feature also appears in the lower panel which shows the rates for BH-BH binaries at different $\kappa$'s.

\begin{figure*}[!]
    \centering
    \includegraphics[width=0.45\textwidth]{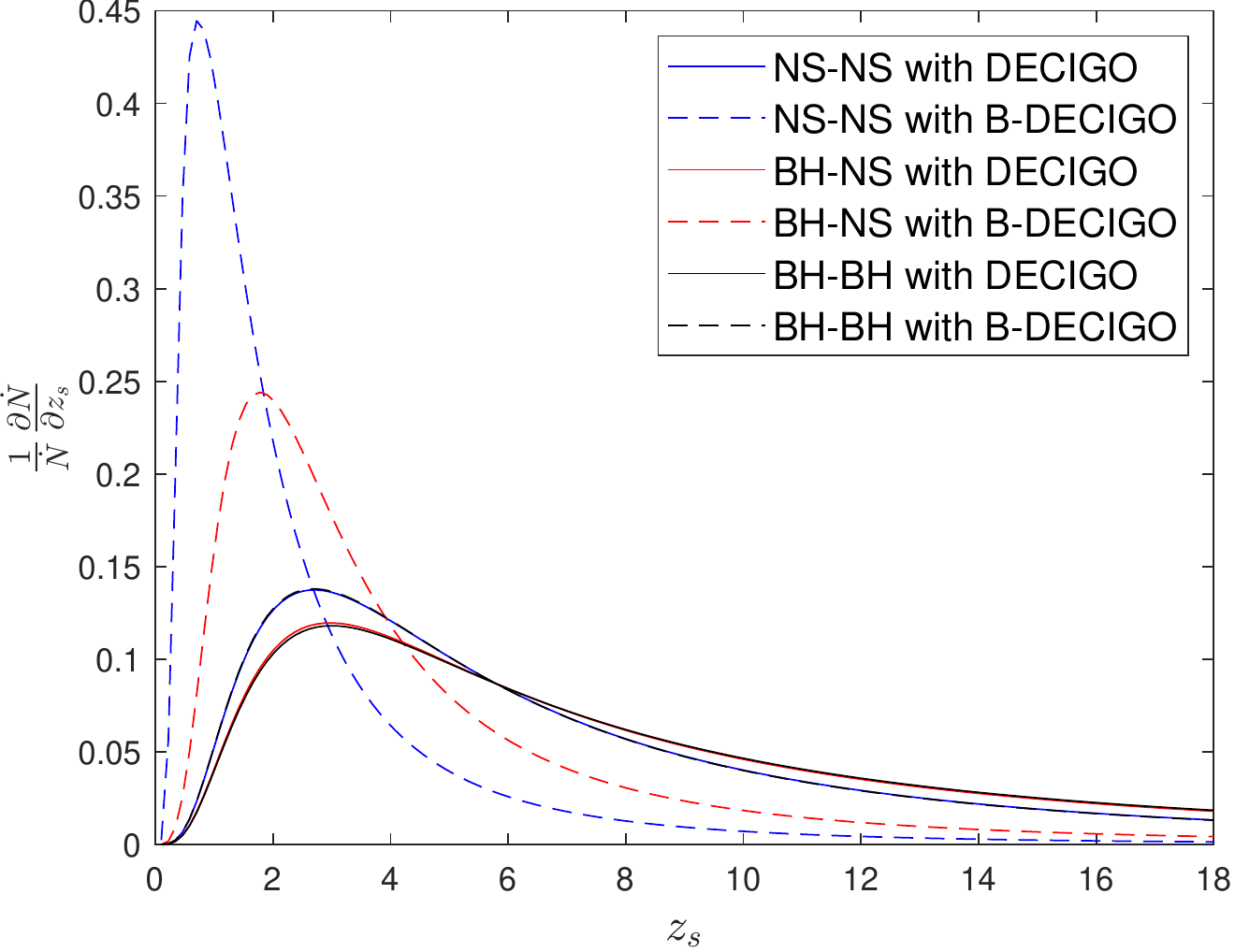}
    \includegraphics[width=0.45\textwidth]{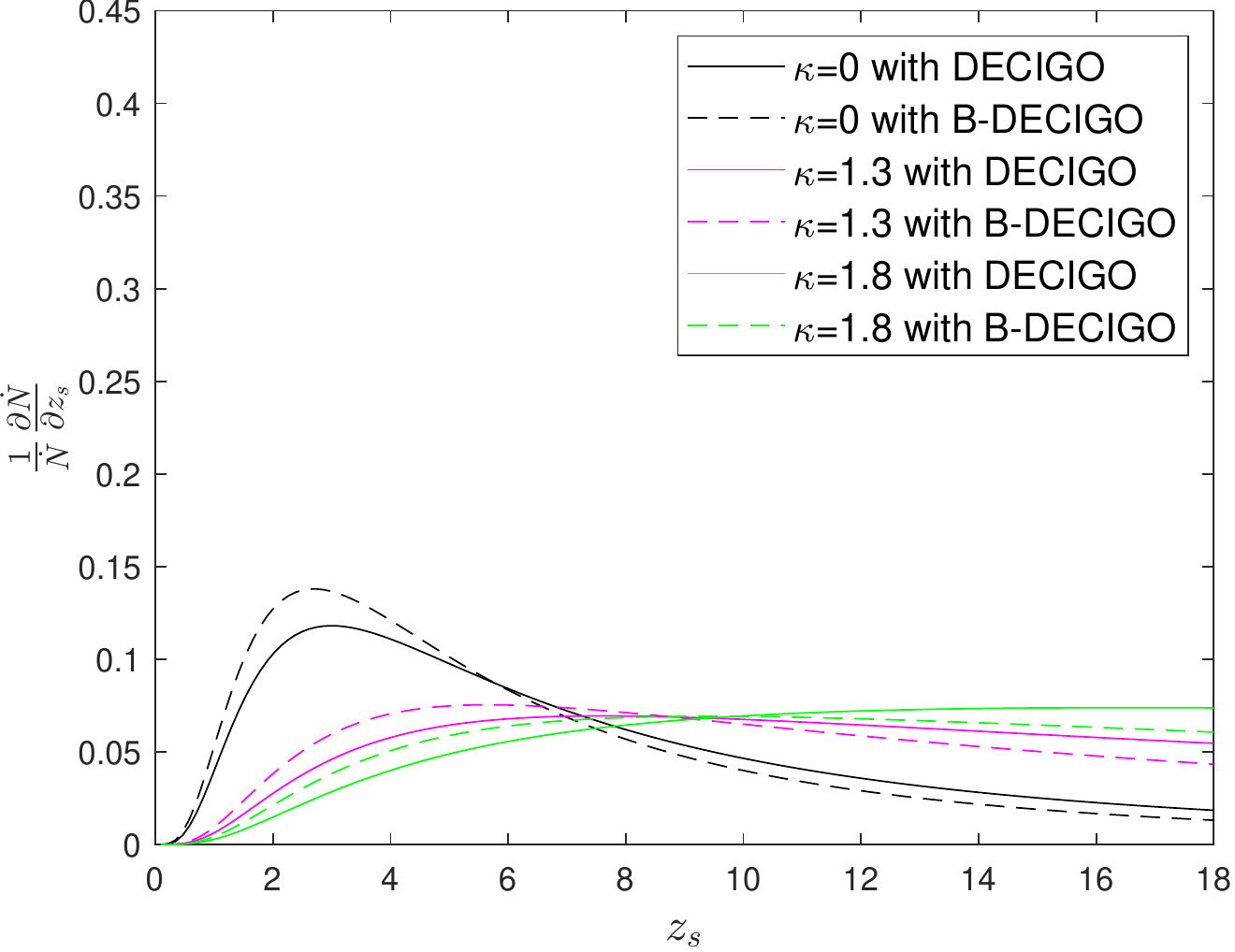}
    \caption{The relative differential detection rates $\frac{1}{\dot N}\frac{\pd\dot N}{\pd z_\text{S}}$ v.s. $z_\text{S}$ for (B-)DECIGO.
    The threshold SNR is 8.
    The left panel shows the rates at $\kappa=0$ for various types of binary systems, and the right panel shows the rates for BH-BH systems at various $\kappa$'s.
    Note in the left panel, the black dashed curvature and the blue solid one nearly overlap each other, so do the red solid and the black solid ones.
    The black curves in the right panel are the same as the black ones in the left panel.}
    \label{fig-app-zs}
\end{figure*}

Figure~\ref{fig-app-zs} displays the relative differential rates $\frac{1}{\dot N}\frac{\pd\dot N}{\pd z_\text{S}}$, independent of $\mathcal R$, too.
The left panel shows the rates for different types of binary systems, assuming the merger rates are independent of the redshift, i.e., $\kappa=0$.
One finds out that B-DECIGO generally detects low redshift sources ($z_\text{S}\lesssim3$), while DECIGO are more sensitive to higher redshift sources ($2<z_\text{S}<4$).
The right panel shows the rates for BH-BH binaries at different $\kappa$'s.
From this, one knows that as $\kappa$ increases, both detectors able to detect more sources at high redshifts.

At $\rho_\text{th}=80$, the lensing rates are definitely smaller, as displayed in Table~\ref{tab-nor80}.
One can see that the capability of DECIGO detecting lensed BH-BH  events remain almost the same, similar to what is found in the previous appendix.
\begin{table}[h]
    \centering
    \bgroup
    \def\arraystretch{1.2}
    \resizebox{\linewidth}{!}{
    \begin{tabular}{c|c|c|c|c|c|c}
        \hline
        \multirow{2}{*}{Binary}& \multicolumn{3}{c|}{DECIGO} & \multicolumn{3}{c}{B-DECIGO} \\
        \cline{2-7}
        & $\kappa=0$ & $\kappa=1.3$ & $\kappa=1.8$& $\kappa=0$ & $\kappa=1.3$ & $\kappa=1.8$\\
        \hline
        NS-NS & 0.002 & - & - & $10^{-7}$ & - & -  \\
        BH-NS & 0.067 & - & - & $2\times10^{-5}$ & - & - \\
        BH-BH & 0.207 & 2.98 & 9.40 & 0.002 & 0.011 & 0.026 \\
        \hline
    \end{tabular}
    }
    \egroup
    \caption{Normalized lensing rates at $\rho_\text{th}=80$.}
    \label{tab-nor80}
\end{table}
The lensing rates at $\rho_\text{th}=800$ are even smaller, as expected, so we will not present them.

\bibliography{apssamp_2.bbl}

\end{document}